%% file: 000paper.tex
\documentclass[]{ceurart}

\usepackage{amsmath,amssymb,amsfonts}
\usepackage{algorithmic}
\usepackage{graphicx}
\usepackage{textcomp}
\usepackage{multirow}
\usepackage{multicol}
\usepackage{hyperref}

\usepackage{color}
\usepackage{xspace}
\usepackage{adjustbox}
\usepackage[dvipsnames]{xcolor}
\usepackage{algorithmic}
\usepackage{colortbl}
\usepackage{tabu}
\usepackage{amsmath}
\usepackage{multibib}
\def\BibTeX{{\rm B\kern-.05em{\sc i\kern-.025em b}\kern-.08em
    T\kern-.1667em\lower.7ex\hbox{E}\kern-.125emX}}

\renewcommand{\baselinestretch}{1.0}

\usepackage{subfig}

\usepackage{tikz}
\usepackage{amsmath}

\usepackage{multicol}
\usepackage{comment}
\usepackage{float}
\usepackage{csquotes}
\usepackage{textcomp}

\usepackage{filecontents}
\usepackage{xspace}
\usepackage{multirow}

\usepackage{multicol}
\usepackage{lipsum}
\usepackage{mwe}

\usepackage{framed}
\usepackage{paralist}
\definecolor{shadecolor}{RGB}{192,192,192}

\usepackage[ruled,linesnumbered]{algorithm2e}

\newcommand{\testAccuracy}{89.43\%\xspace}
\newcommand{\valAccuracy}{89.44\%\xspace}
\newcommand{\trainAccuracy}{95.95\%\xspace}
\usepackage{listings}
\usepackage{float} 
\usepackage{placeins} 

\clearpairofpagestyles 
\ofoot{\thepage} 

\begin{document}
\copyrightyear{2024}
\copyrightclause{Copyright for this paper by its authors.
  Use permitted under Creative Commons License Attribution 4.0
  International (CC BY 4.0).}
\conference{eCom'24: ACM SIGIR Workshop on eCommerce, July 18, 2024, Washington, DC, USA}

\title{Enhancement of E-commerce Sponsored Search Relevancy with LLM}


\author[1]{Md Omar Faruk Rokon}[%
email=mdomarfaruk.rokon@walmart.com,
]
\cormark[1]
\address[1]{Walmart AdTech, Sunnyvale, CA, USA}

\author[1]{Andrei Simion}[%
email=andrei.simion@walmart.com,
]

\author[1]{Weizhi Du}[%
email=weizhi.du@walmart.com,
]

\author[1]{Musen Wen}[%
email=musen.wen@walmart.com,
]

\author[1]{Hong Yao}[%
email=hong.yao0@walmart.com,
]

\author[1]{Kuang-chih Lee}[%
email=kuang-chih.lee@walmart.com,
]
  
\cortext[1]{Corresponding author.}

\begin{keywords}
  LLM \sep
  LLaMa \sep
  LoRA \sep
  Relevance \sep
  Sponsored Search
\end{keywords}

\maketitle

\input{001abstract}

\section{Introduction}
\label{sec:intro}
\input{010introduction}

\section{Related Works}
\label{sec:related_works}
\input{050related}

\section{Our Approach}
\label{sec:meth}
\input{030method}


\section{Experimental Results}
\label{sec:results}
\input{041results}

\section{Applications of the Relevance Model}
\label{sec:applications}
\input{042application}

\section{Discussion}
\label{sec:discuss}
\input{045discussion}

\section{Conclusions}
\label{sec:concl}
\input{060conclusion}

\bibliography{BIB/rokon}

\end{document}

%% file: 001abstract.tex
\begin{abstract}
Sponsored search plays a crucial role as a revenue stream for search engines, wherein advertisers competitively bid on keywords that align with the users' search queries. The task of matching relevant keywords to these queries is complicated by the vast and ever-evolving space of keywords, the ambiguity of user and advertiser intentions, and the wide range of topics and languages involved. Consequently, ensuring that ads are pertinent to user queries presents significant challenges.
In the fast-paced world of e-commerce, the accuracy of sponsored search results is vital for boosting user satisfaction and optimizing business operations. This paper presents the development of an advanced Ad Relevance Model within a sponsored search framework, utilizing the power of a pretrained large language model. We detail a pioneering adaptation of the LLAMA2 7B model through Low-Rank Adaptation (LoRA), which markedly enhances search precision and operational efficiency, thus opening new avenues for improving user interactions in extensive online marketplaces such as Walmart.com.
We introduce a novel $<$query, ad title$>$ classifier, which discerns the relevance of search interactions across three categories: Relevant, Partially Relevant, and Irrelevant. Our approach involved adapting the pretrained model specifically for the e-commerce sponsored search context, training it on a large dataset. The fine-tuned model demonstrated a marked improvement in ad relevance accuracy, achieving \testAccuracy accuracy on a comprehensive test dataset—outperforming both the baseline model and other advanced language models like GPT-4. The integration of LoRA with the based model represents a significant stride in customizing language models for e-commerce applications, resulting in enhanced search accuracy, cost efficiency, and operational privacy—a triad essential for the modern digital marketplace.

\end{abstract}


%% file: 010introduction.tex
The advent of e-commerce has revolutionized the retail landscape, creating a pressing need for advanced technological solutions to improve the user experience. A pivotal aspect of this experience is the relevance of sponsored product searches—a factor that directly influences customer satisfaction and retention\cite{aiello2016role, abhishek2019advertising, Wang2023ClickConversion}. Traditionally, search relevance in e-commerce platforms has been tackled using various algorithmic approaches, but these often fall short in understanding the nuanced language of consumer queries \cite{rao2020product,su2018user, Li2023CommunicativeMARL}.

Recent advances in Natural Language Processing (NLP), particularly the advent of large language models (LLMs), have brought a revolution in solving complex information retrieval problems. These models have significantly improved our ability to interpret and respond to the intricate nuances of human language, presenting new opportunities for enhancing search relevance. Despite this progress, there remains a significant gap in the application of these models within the unique constraints of e-commerce search environments, where the interpretation must be quick, accurate, and commercially viable.

The importance of search relevance in e-commerce cannot be overstated. A customer's experience, and consequently their purchasing decisions, hinge on the ability of search engines to understand and accurately respond to their queries. For instance, a query as brief as ``waterproof hiking boots" requires the system to interpret not just the literal request but also the implied preferences such as durability, brand preferences, and price range. However, traditional techniques often fail to grasp these subtleties, leading to sub-optimal search results and, in turn, affecting crucial business metrics like click-through and conversion rates.

Our research is motivated by these challenges and the potential of LLMs to revolutionize search relevance in e-commerce. We propose the following research question: How can the advanced capabilities of LLMs, specifically in understanding brief and complex query intents, be effectively harnessed to enhance the relevance of sponsored product searches in e-commerce platforms, overcoming the limitations of traditional search algorithms?

Addressing this, we employed LLAMA2 7B, a decoder based model trained on vast natural language for its human-like text comprehension and generation abilities. We adapted it using Low-Rank Adaptation (LoRA) which allows for efficient model fine-tuning without the need for extensive computational resources \cite{hu2021lora}.

\begin{figure*}[ht]
\centering
\includegraphics[width=0.8\textwidth]{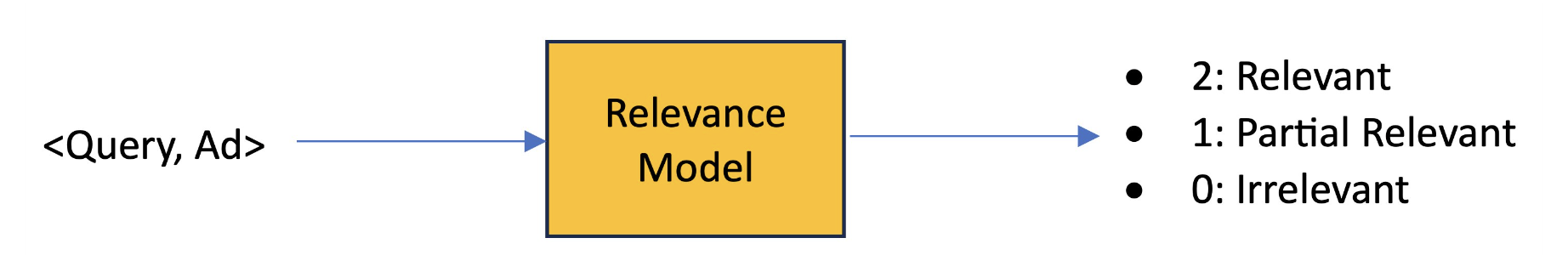}
\caption{Given a query and Ad item pair, the Relevance Model predicts their relevance score, categorizing them as `Relevant', `Partially Relevant', or `Irrelevant'.}
\label{fig:problem}
\end{figure*}

As illustrated in Figure \ref{fig:problem}, our approach to solving these challenges involves developing a $<$query, ad title$>$ classification system. This system categorizes the relevance of search interactions more precisely, assigning a relevance score that drives the decision-making process for ad placement. Furthermore, we present a comprehensive comparison of our fine-tuned LLAMA2 7B model against existing models, including GPT-4, to demonstrate its superior performance in terms of accuracy, efficiency, and cost-effectiveness.


Our contributions are as follows:
\begin{itemize}
\item Demonstration of the inherent ability of the llama model to understand the intents behind e-commerce queries such as ``energy-efficient LED bulbs" and ``compact high-resolution cameras." This capability marks a significant advancement in interpreting complex query intents and item texts, leading to more relevant and tailored search results without the need for a separate query understanding system.
\item Enhancement of search relevance accuracy through a specialized classification system, improving user experience and business metrics.
\item We develop an advanced system for auto-labeling query and product relevance, facilitating continuous model training.
\item Comparative analysis showcasing the advancement of our model over traditional NLP models in the e-commerce domain.
\end{itemize}

This paper provides a detailed account of our methods, the challenges overcome, and the implications of our findings for the future of e-commerce search relevance. By sharing our approach and results, we aim to offer a valuable reference point for future research and development in the field of NLP as applied to e-commerce search engines.

While we evaluate this model specifically for sponsored search, the approach is equally applicable to broader e-commerce search contexts. This extension underscores the versatility and practical utility of our method across various e-commerce search scenarios.

%% file: 050related.tex
This section provides a comprehensive review of the literature in e-commerce search relevance, the role of large language models (LLMs) in NLP, the application of Low-Rank Adaptation (LoRA) techniques, and the specific use of the LLAMA2 7B model. Our aim is to contextualize our research within the broader scientific discourse, highlighting the innovative aspects of our approach.

\subsection{E-commerce Search Relevance}
E-commerce search relevance has been a critical area of research, focusing on improving the accuracy and user experience of search engines. Traditional algorithms such as keyword matching and collaborative filtering have laid the foundation for early systems \cite{Rao2020}. However, the dynamic nature of user queries and the diversity of inventory make it challenging to maintain high relevance \cite{Su2018}. The impact of search relevance on business metrics, including click-through rates and conversion rates, underscores its importance for e-commerce platforms \cite{Aiello2016}.

\subsection{Large Language Models in NLP}
The development and application of LLMs have significantly advanced the field of NLP. Transformer-based models, like GPT-4, have set new standards for understanding and generating human-like text \cite{achiam2023gpt}. These models' adaptability for domain-specific tasks highlights their potential beyond generic NLP applications \cite{hu2021lora}. The LLAMA2 by Meta \cite{touvron2023llama} represents a significant leap in the application of LLMs for e-commerce search relevance. Its architecture and training methodologies are specifically designed to cater to the complex requirements of understanding consumer queries \cite{Wang2023}. Comparisons with other models demonstrate LLAMA2 7B's superior performance in text understanding and generation tasks, and the model being the opensource model, making it an ideal candidate for enhancing e-commerce search systems.

\subsection{Low-Rank Adaptation (LoRA)}
The concept of LoRA has emerged as an efficient technique for adapting large models without extensive computational resources. By introducing low-rank matrices, LoRA allows for the selective fine-tuning of model parameters, maintaining high performance while reducing the computational footprint \cite{hu2021lora}. Comparative studies have illustrated LoRA's efficacy against other model adaptation techniques, emphasizing its utility in various contexts \cite{Li2023}.

In summary, our research leverages the strengths of the LLAMA2 7B model, enhanced through LoRA, to address the nuanced challenges of e-commerce search relevance. By integrating these advanced NLP techniques, we aim to set a new benchmark for search accuracy and efficiency.

%% file: 030method.tex
We fine-tuned the model for efficient adaptation to our specific e-commerce context to enhance the relevance of sponsored product searches on Walmart.com. Our solution employs the integration of Low-Rank Adaptation (LoRA) with the LLAMA2 7B model, optimizing resource utilization and performance.

\subsection{System Architecture}

The architecture of our model is a cornerstone of our methodology, involving the parameter-efficient fine-tuning model (Peft Model) incorporating the LLAMA2 7B model, which was fine-tuned using the Low-Rank Adaptation (LoRA) technique. This approach allows us to maintain computational efficiency while significantly enhancing model performance specific to e-commerce search queries.

\textbf{Low-Rank Adaptation (LoRA)}: LoRA is a technique designed to efficiently adapt large pre-trained models like LLAMA2 7B to specific tasks without the need for extensive retraining, thus preserving computational resources while improving performance on specialized tasks. Given a weight matrix \( W \) of dimensions \( d \times h \), LoRA introduces two low-rank matrices \( A \) and \( B \) of dimensions \( d \times r \) and \( r \times h \) respectively, where \( r \) is significantly smaller than \( h \). The adapted weight matrix \( \hat{W} \) is then computed as \( W + AB \). This adaptation is performed while keeping \( W \) frozen during training, and only \( A \) and \( B \) are updated, leading to a more parameter-efficient model adaptation. The effectiveness of LoRA can be quantified through performance improvements in model-specific tasks.

In operation, the model processes input text through its embedding layers, followed by sequential passage through the decoder layers of the LLAMA model, each applying self-attention and feed-forward mechanisms now fine-tuned with LoRA, to effectively understand and categorize the textual input relevant to e-commerce search queries. The final classification is made by the linear score module at the end of the network. This section details the structure of our model, emphasizing the key components that contribute to its performance in sponsored product search relevance.

\begin{figure*}[ht]
\centering
\includegraphics[width=0.7\textwidth]{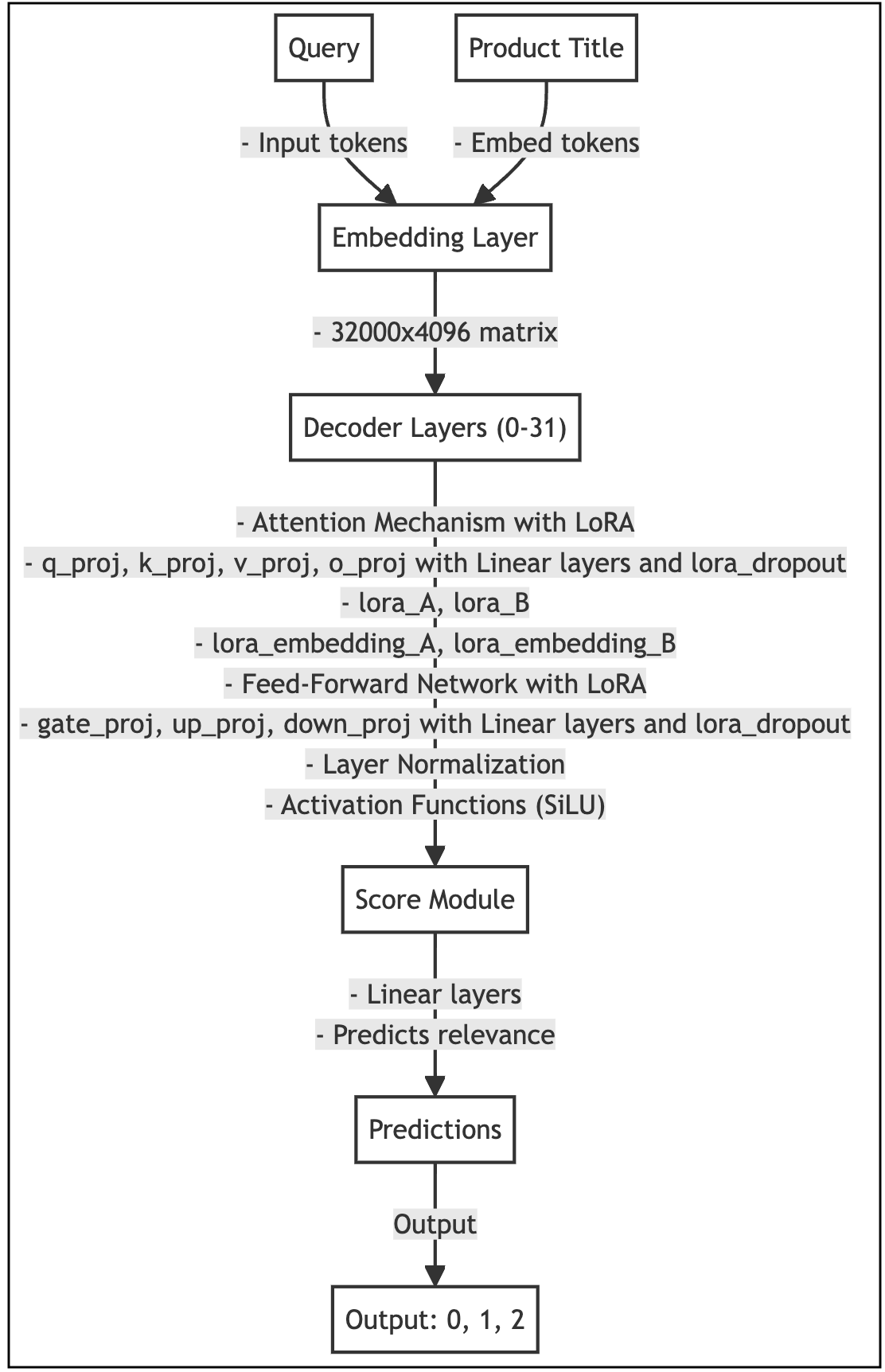}
\caption{Overview of our model fine-tuning using LoRA}
\label{fig:model_architecture}
\end{figure*}


\textbf{Model Overview:} The base of our model is the LLAMA2 7B, a large language model known for its robust text understanding and generation capabilities. It is structured around a decoder layer of the LLaMa model, which is responsible for processing the input sequences. The core components of the model include:

\begin{itemize}
    \item \textbf{Embedding Layers}: Responsible for converting input tokens into dense vectors. It includes an embedding matrix of size 32000x4096.
    \item \textbf{Decoder Layers}: A total of 32 layers, each containing components for attention mechanisms and feed-forward networks.
\end{itemize}

\textbf{Low-Rank Adaptation Components:} LoRA is applied to key components of the LLAMA2 7B model, enabling efficient and effective fine-tuning. The adapted components include:

\begin{itemize}
    \item \textbf{Attention Mechanism}: Adaptation is applied to the query (q\_proj), key (k\_proj), value (v\_proj), and output (o\_proj) projection layers of the attention mechanism.
    \item \textbf{Feed-Forward Networks}: Includes adaptations to the gate projection (gate\_proj), up projection (up\_proj), and down projection (down\_proj) layers in the feed-forward networks.
\end{itemize}

Each adapted component consists of a set of Linear layers and Dropout modules, with LoRA introducing additional Linear layers (lora\_A and lora\_B) and parameter matrices (lora\_embedding\_A and lora\_embedding\_B) for each component.

\textbf{Normalization and Activation Functions: } Layer normalization (LlamaRMSNorm) and activation functions (SiLUActivation) ensure stable training and introduce non-linearity to the model's operations.

This architecture enables our model to efficiently learn and adapt to the specific nuances of e-commerce search queries, thus enhancing the relevance of search results. The use of LoRA ensures that this adaptation is both computationally efficient and effective, maintaining the robustness of the LLAMA2 7B model while tailoring it to our specific use case.

\subsection{Training}
We fine-tune the model using a large dataset specific to the e-commerce domain, enhancing the model's capability to discern nuanced e-commerce search queries without incurring significant computational costs.


\textbf{Model Adaptation with LoRA:}
We adapt the LLAMA2 7B model using the Low-Rank Adaptation (LoRA) technique, as described earlier. This adaptation method is crucial because it effectively reduces the number of trainable parameters, thereby enhancing parameter efficiency during model tuning. The primary reason for applying LoRA is its ability to allow significant modifications to the model's behavior with minimal updates to its parameters. This is particularly valuable in scenarios like ours, where computational resources are limited and retraining large models from scratch is computationally expensive and time-consuming. By applying LoRA, we ensure that the model remains lightweight and agile, capable of adapting to the nuances of e-commerce search queries without the need for extensive computational power. This approach not only conserves resources but also accelerates the deployment cycle, making it highly suitable for dynamic e-commerce environments where prompt model updates are often required.

\textbf{Training Setup:}
Our training involves a dataset where each example comprises a text query paired with a relevance label. The model is configured to classify these examples into three relevance categories: Irrelevant, Partially Relevant, and Relevant. Training is conducted over several epochs with a batch size of 8. For detailed information about the training dataset, please refer to Section \ref{subsec:dataset}.

\textbf{Optimization and Loss Function:} 
The optimization of the model is a critical step in the training process. We define the objective function \(\mathcal{L}\) for our multi-class classification task as the categorical cross-entropy loss, given by:

\begin{equation}
\mathcal{L}(\theta) = - \frac{1}{N} \sum_{i=1}^{N} \sum_{c=1}^{C} y_{i,c} \log(p_{i,c}(\theta))
\end{equation}

where \(N\) is the number of examples, \(C\) is the number of classes, \(y_{i,c}\) is a binary indicator of whether class label \(c\) is the correct classification for observation \(i\), and \(p_{i,c}(\theta)\) is the predicted probability that observation \(i\) is of class \(c\), parameterized by \(\theta\).

The model parameters are optimized using the Adam optimizer, a stochastic gradient descent method with an adaptive learning rate, described by:

\begin{equation}
\theta_{t+1} = \theta_{t} - \frac{\eta}{\sqrt{\hat{v}_t} + \epsilon} \hat{m}_t
\end{equation}

where \(\theta_{t}\) represents the parameters at iteration \(t\), \(\eta\) is the step size, \(\hat{m}_t\) and \(\hat{v}_t\) are estimates of the first and second moments of the gradients, and \(\epsilon\) is a small scalar used to prevent division by zero.

Utilizing an optimizer like Adam and a Cross Entropy loss function, we ensured efficient learning during the training process.
This loss function and optimization algorithm were iteratively applied to update model parameters and minimize the objective function over epochs.




The design and implementation of this solution aimed to balance computational efficiency with the requirement for high accuracy in classifying the relevance of search queries, thus addressing the core challenge in sponsored product search relevance for e-commerce platforms.

%% file: 041results.tex
Our experimental framework aimed to rigorously evaluate the performance of the fine-tuned LLAMA2 7B model with LoRA in the context of sponsored product search relevance. The experiments were conducted using datasets representative of real-world e-commerce search queries and their corresponding relevance classifications.

The training and evaluation of our model were conducted using a high-performance setup featuring 4 Tesla V100-SXM2-32GB GPU. This hardware configuration played a crucial role in managing the computational demands, ensuring efficient and effective training and evaluation of our model using LoRA.

\subsection{Dataset}
\label{subsec:dataset}
Our experimental framework utilizes a comprehensive dataset derived from Walmart's query logs, comprising 250K query and item title pairs for training, alongside 56K pairs each for validation and testing. These query item pairs (QIPs) were sampled using a stratified approach to ensure a balanced representation of relevant, partially relevant, and irrelevant classes.  Third-party annotators meticulously evaluated these QIPs to assess relevancy on a three-point scale: 0 for irrelevant, 1 for partially relevant, and 2 for highly relevant. Given the subjective nature of relevancy, each pair was scored by three independent annotators, with the majority vote determining the final relevance score. This rigorous annotation process ensures a high-quality dataset, crucial for the accurate training and evaluation of our model.

\begin{table}[h]
\centering
\begin{tabular}{lcccc}
\hline
Dataset & Irrelevant & Partially Relevant & Relevant & Total \\
\hline
Training & 98,938 (39.6\%) & 84,388 (33.8\%) & 66,229 (26.5\%) & 250K \\
Validation & 16,252 (29.0\%) & 21,603 (38.6\%) & 17,644 (31.5\%) & 56K \\
Testing & 16,281 (29.1\%) & 21,616 (38.6\%) & 17,598 (31.4\%) & 56K \\
\hline
\end{tabular}
\caption{Distribution of query-item pairs in the dataset}
\label{table:dataset_distribution}
\end{table}

The distribution of query-item pairs across training, validation, and testing sets is summarized in Table \ref{table:dataset_distribution}. The varied distribution reflects the real-world complexity of e-commerce search relevance, with a notable class imbalance that poses a challenge for accurate classification. This underscores the need for a model that not only achieves high accuracy but also maintains precision across diverse relevance categories, a critical factor for the success of search relevance algorithms in e-commerce.

\subsection{Training and Validation}

\begin{figure}[h]
\centering
\includegraphics[width=0.8\linewidth]{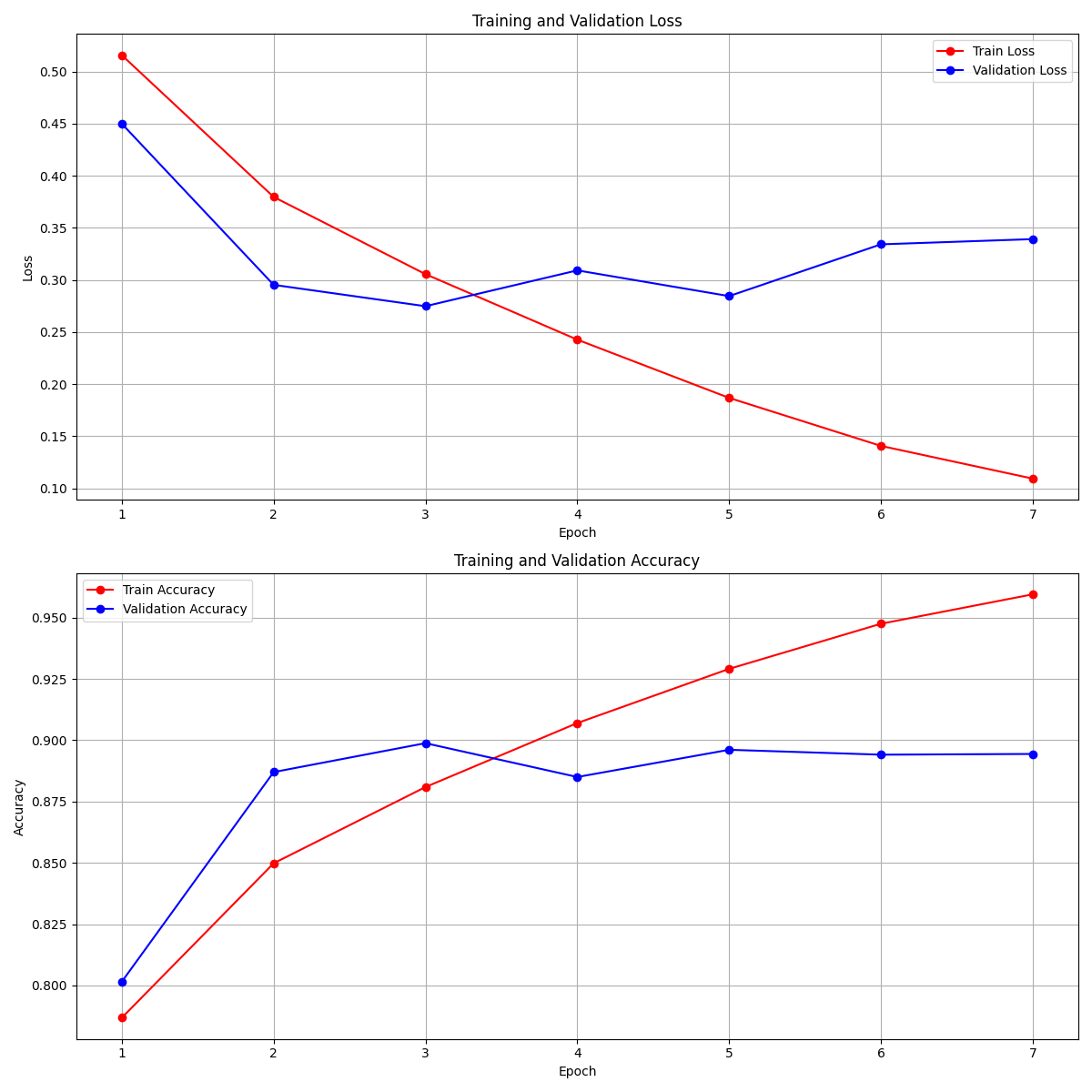}
\caption{Validation Performance During Training: This figure illustrates the model's loss and accuracy on the validation dataset over training epochs, demonstrating the learning progress and convergence behavior.}

\label{fig:validation_performance}
\end{figure}

The model's training was conducted over 7 epochs with a batch size of 16. We employed the Adam optimizer, setting the learning rate to 1e-5, and used Cross-Entropy Loss to guide the optimization process. The training process is visualized in Figure \ref{fig:validation_performance}, which depicts the loss and accuracy curves over the training epochs. Initially, we observe a rapid decline in loss, indicating that the model is quickly learning from the training data. The accuracy, correspondingly, increases sharply and then plateaus, suggesting that the model is achieving a stable understanding of the data. This plateau may imply that the model has reached its capacity for learning from the provided data, or it could indicate the need for a more complex model or additional features to capture the underlying patterns better. 
At the end of the training, we selected the epoch 3 model as the best performing model which got \trainAccuracy as the training accuracy and \valAccuracy as the validation accuracy.

\begin{table}[h]
\centering
\begin{tabular}{lcccc}
\hline
Metric & Training & Validation & Testing \\
\hline
Accuracy & 95.95\% & 89.44\% & 89.43\% \\
Precision & N/A & 89.75\% & 89.75\% \\
Recall & N/A & 89.23\% & 89.22\% \\
F1-Score & N/A & 89.4\% & 89.41\% \\
\hline
\end{tabular}
\caption{Performance metrics of the fine-tuned LLaMa2 7B model on the test dataset}
\label{table:model_performance}
\end{table}


\subsection{Evaluation on Test Set}
The model's generalization capabilities were rigorously assessed on an independent test set, the results of which are consolidated in Table \ref{table:model_performance}. The favorable metrics underscore the model's adeptness at maintaining high performance on unseen data, a testament to its robustness. Particularly, the F1-Score—a balanced measure of precision and recall—suggests that the model is well-tuned across all categories of search relevance, an essential attribute for real-world application where diverse query intents are prevalent.

\textbf{Confusion Matrix Analysis: }
The confusion matrix provided insights into the model's performance across different relevance classes. 

\begin{figure}[h]
\centering
\includegraphics[width=0.8\linewidth]{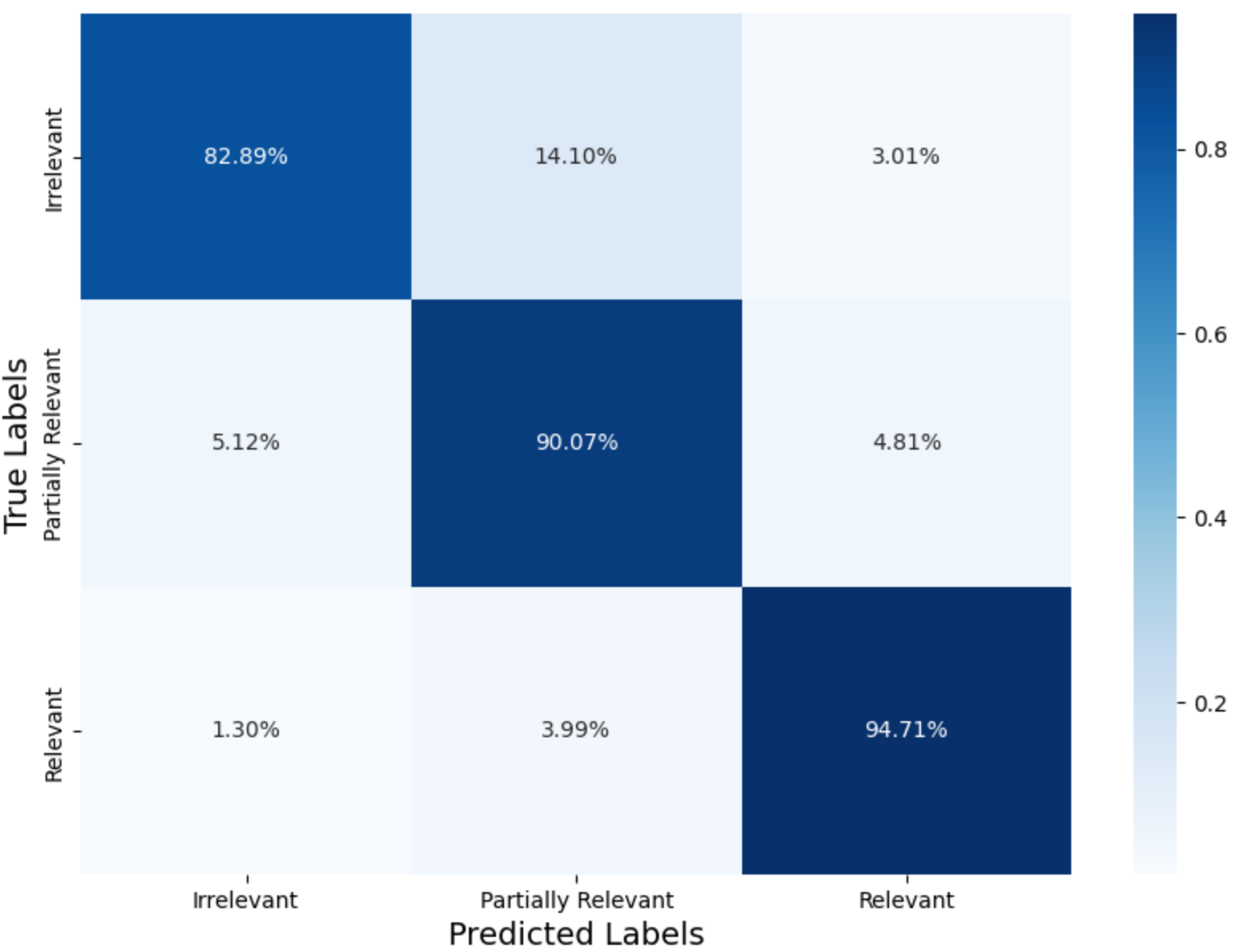}
\caption{Test Data Confusion Matrix: Percentage view of the model’s classification performance across different relevance categories on the test dataset, highlighting its ability to distinguish between irrelevant, partially relevant, and relevant queries.}
\label{fig:confusion_matrix_percentage}
\end{figure}

In Figure \ref{fig:confusion_matrix_percentage}, the confusion matrix offers insights into the model's predictive capabilities across the three categories of search query relevance. Notably, the model demonstrates high precision in identifying `Relevant' query item pairs, with a 94.71\% correct classification rate, which is critical for user satisfaction in e-commerce search scenarios. The `Partially Relevant'  category also shows strong model performance with 90.07\% accuracy. However, the model appears to struggle more with `Irrelevant'  query item pairs, as indicated by the off-diagonal percentages in this row, suggesting a potential area for model refinement.


\subsection{Comparative Evaluation}
In this section, we conduct a comparative evaluation of our fine-tuned LLAMA2 7B model against both traditional and state-of-the-art models previously utilized for semantic retrieval and relevance in sponsored product searches at Walmart. Specifically, we compare our model to a BERT-based Bi-Encoder, a BERT-based Cross-Encoder, and GPT-4 using a few-shot learning approach.

\textbf{Evaluation Setup:} 
We evaluate the models in two key areas:

{\em Offline Performance:} Accuracy, precision, recall, and F1-score metrics are calculated using the same test dataset to ensure a fair comparison.

{\em Relevance Analysis:} We assess model relevance using Walmart Sponsored Search data. For this analysis, we prepare a dataset of 1,000 queries where the top 20 items are retrieved based on the models' scores. These items are then evaluated for their relevancy by third-party human annotators, creating a human-labeled dataset of 30,000 query-item pairs.

\textbf{Offline Metrics:}
The table below presents a side-by-side comparison of the models based on their performance metrics. Our fine-tuned LLAMA2 7B model demonstrates a superior balance between precision and recall, indicating a more nuanced understanding of query-item relevance, which is essential for reducing both false positives and false negatives in e-commerce settings.

\begin{table}[h]
\centering
\begin{tabular}{ccccc}
\hline
Model & Accuracy & Precision & Recall & F1-score \\
\hline
Fine-tuned LLaMa2 7B & 89.43\% & 89.75\% & 89.22\% & 89.41\% \\
Cross-Encoder BERT & 86.27\% & 86.83\% & 86.58\% & 86.70\% \\
Bi-Encoder BERT & 74.42\% & 75.65\% & 94.09\% & 83.87\% \\
GPT-4 Few-shot & 63.02\% & 66.40\% & 61.93\% & 62.88\% \\
\hline
\end{tabular}
\caption{Performance comparison of fine-tuned LLAMA2 7B, Bi-Encoder BERT, and GPT-4 models}
\label{table:performance_comparison}
\end{table}



The fine-tuned LLAMA2 7B model not only achieves the highest accuracy at 89.43\%, but also maintains a balanced precision (89.75\%) and recall (89.22\%), leading to an F1-score of 89.41\%. This balance is crucial for ensuring that relevant items are identified without overwhelming users with irrelevant results. In contrast, GPT-4 performs significantly worse, with the lowest accuracy at 63.02\%, precision at 66.40\%, recall at 61.93\%, and F1-score at 62.88\%. This indicates that GPT-4 struggles with both precision and recall, making it less suitable for high-stakes e-commerce environments where accuracy and relevance are critical. The Cross-Encoder BERT shows respectable performance but still falls short of LLAMA2 7B, while the Bi-Encoder BERT, despite its high recall, suffers from lower precision, highlighting its tendency to over-retrieve irrelevant items.

These findings underscore the importance of model selection in search relevance tasks and pave the way for future research on optimizing retrieval models in large-scale e-commerce platforms. The superior performance of the fine-tuned LLAMA2 7B model demonstrates its potential to significantly improve user experience and business efficiency in online retail environments.

\textbf{Relevance Analysis:}
We further assess the relevancy of the returned ad items to the 1000 user queries using the Normalized Discounted Cumulative Gain (NDCG) metric at cutoffs of 4 and 8. This metric measures the usefulness, or gain, of the ad items based on their positions in the result list, with higher scores indicating better relevancy. In Table \ref{table
}, we show the relevance comparison of the different models. We did not include GPT-4 in this analysis due to the significant performance gap observed in the offline metrics and to avoid the complexity and cost associated with its API usage.

\begin{table}[h]
\centering
\begin{tabular}{ccccc}
\hline
Model & NDCG@4 & NDCG@8 \\
\hline
Fine-tuned LLaMa2 7B & 0.7142 & 0.6774 \\
Cross-Encoder BERT & 0.6939 & 0.6565 \\
Bi-Encoder BERT & 0.6685 & 0.6324 \\
\hline
\end{tabular}
\caption{Relevance comparison of fine-tuned LLAMA2 7B, Cross-Encoder BERT, and Bi-Encoder BERT}
\label{table:ndcg_comparison}
\end{table}

The experimental results demonstrate that the fine-tuned LLAMA2 7B model, augmented with LoRA, not only maintains high performance on unseen data but also significantly enhances the accuracy of sponsored product search relevance on Walmart.com. Its robust performance across both validation and testing phases underscores its practical applicability in a real-world e-commerce setting. The model's ability to effectively classify queries into relevant categories illustrates its potential to substantially improve user experience and business efficiency in online retail platforms, highlighting its robustness and the real-world utility in e-commerce search environments.

%% file: 042application.tex
The relevance model plays a critical role in enhancing the performance and efficiency of sponsored ads systems. In the context of search pages, profitability arises from both the payments received from sellers for clicks on their ads and from improved purchase conversion rates when the relevance of ads is high. By accurately determining the relevance of ads to user queries, the model not only improves user experience but also maximizes the profitability of ad placements. This section outlines key applications of the relevance model in both current and future sponsored ads systems.

\textbf{Improving Relevancy in Current Systems: }
Current applications of the relevance model focus on enhancing the precision and accuracy of ad targeting. Immediate steps include:

{\em 1. Quick Evaluation of Current System Relevance:} Regular assessments of the model's performance are crucial. By analyzing current system outputs, particularly the relevance of ads served to user queries, we identify immediate areas for improvement.

{\em 2. Identification of Relevance Issues and Gaps:} This involves a thorough examination of the training data and serving logs to pinpoint discrepancies that may lead to suboptimal model performance. Identifying these gaps allows for targeted data corrections and system adjustments.

{\em 3. Iterative Improvement Based on Feedback:} Utilizing a feedback loop from system outputs and user interactions, the model undergoes continual refinements. This iterative process ensures that the model adapts to changing user behaviors and market trends, maintaining its relevance and effectiveness.

\textbf{Enhancing Relevancy for Future Systems: }
Looking forward, the relevance model will incorporate advanced technologies and methodologies to further refine ad targeting mechanisms:

{\em 1. Evaluation System Enhancements:} Future systems will focus on increasing modeling productivity through innovations such as query rewriting and the integration of multimodal models. These advancements aim to understand and process user intents more comprehensively.

{\em 2. Automated Labeling Systems:} Implementing auto-labeling technologies promises significant reductions in both time and cost associated with manual data labeling. This automation will facilitate the rapid scaling of training datasets, enhancing the model's learning capabilities and accuracy over time. However, given the issue with irrelevant recall, some ``partially relevant" items should be manually labeled to double-check accuracy.


\textbf{Cost and Time Efficiency: }
The integration of advanced relevance models and automated systems in sponsored ads not only enhances accuracy and user experience but also brings substantial cost and time savings. These efficiencies manifest in several key areas:

{\em Reduction in Manual Labeling Costs:} By implementing auto-labeling technologies, the need for extensive manual labeling is greatly reduced. Manual data labeling is not only time-consuming but also costly, often requiring the hiring of third-party evaluators to ensure quality and unbiased data. Automated systems streamline this process, significantly decreasing the operational costs associated with data preparation.

{\em Decreased Reliance on Third-Party Evaluators:} Auto-labeling and enhanced model accuracy diminish the dependency on external evaluators. While third-party expertise is invaluable, particularly for validating model predictions and annotating complex datasets, reducing reliance on these services cuts down expenses and accelerates the data preparation phase.

{\em Faster Iteration of Model Development:} Automated and refined processes enable more rapid model iterations. With reduced delays in data preparation and evaluation, models can be updated and refined more quickly. This faster iteration cycle not only saves time but also allows for agile responses to market changes and user feedback, ensuring the system remains at the forefront of relevancy and effectiveness.

Overall, the strategic application of relevance models and automation in the sponsored ads system represents a smart investment. These technologies not only improve the operational aspects of ad placement but also drive substantial economic benefits by optimizing resource allocation and reducing the time to market for system enhancements. This financial and temporal efficiency is crucial for maintaining a competitive edge in the dynamic landscape of digital advertising.

%% file: 045discussion.tex
In navigating the complex landscape of e-commerce search relevance, the intersection of user intent with accurate product matching stands as a cornerstone for enhancing user experience and business outcomes. This section delves into the nuanced interplay of predictive accuracy and user-centered search results, illuminated by our study's examination of advanced language models such as BERT, GPT-4 and LLAMA2. Herein, we discuss the implications of our findings, drawing connections between technological innovation and its translation to practical e-commerce solutions.

\begin{table}[]
\begin{tabular}{|p{2cm}|p{8cm}|p{1cm}|p{1cm}|p{1cm}|}
\hline
\textbf{Query} & \textbf{Item Title} & \textbf{Human Label} & \textbf{GPT-4 Label} & \textbf{LLaMa2 Label} \\ \hline
toilet paper & Charmin Ultra Soft Toilet Paper 6 Mega Rolls, 224 Sheets per Roll & 2 & 2 & 2 \\ \hline
toilet paper & SUGARDAY Toilet Bowl Brush and Caddy Holder Set for Bathroom Soft Bristle Silicone Plunger Set Toilet Scrubber with Tweezers White & 1 & 1 & 1 \\ \hline
tv for living room & SinCiDo Farmhouse TV Stand for TVs Up to 80 inches, 39" Tall Highboy Entertainment Center w/Barn Door, Large Wood Rustic TV Console Cabinet w/Adjustable Shelves for Living Room, 70inch, Dark Grey & 1 & 2 & 1 \\ \hline 
tv for living room & VIZIO 65" Class V-Series 4K UHD LED Smart TV V655-J09 & 2 & 2 & 2 \\ \hline
first aid tape & McKesson Medical Tape 1" x 10 yd 16-47210, 12 Boxes, 12 Rolls/Box & 2 & 1 & 2 \\ \hline
subwoofer & Yamaha NS-IC800 8" In-Ceiling Speaker (Pair, White) & 2 & 0 & 2 \\ \hline
water can & AHA Sparkling Water, Orange + Grapefruit Flavored Water, Zero Calories, Sodium Free, No Sweeteners, 12 fl oz, 8 Pack & 2 & 1 & 2 \\ \hline
shredded chicken & Chicken Shredder Meat Shred Machine for Pulled Pork, Beef and Chicken & 0 & 2 & 0 \\ \hline 
glass cleaner & ZEISS Anti-Fog Lens Wipes, Pre-Moistened Eye Glass Cleaner Wipes, 30 Count & 0 & 2 & 0 \\ \hline 
blueberry & Thomas' Blueberry Pre-Sliced Bagels, 6-Count & 0 & 2 & 0 \\ \hline 
projector & Canon REALiS WX450ST-D Projector Lamp with Module & 0 & 2 & 0 \\ \hline 
dairy free cheese & Daiya Dairy Free Gluten Free Cheese Lover's Vegan Pizza, 15.7 oz (Frozen) & 0 & 2 & 0 \\ \hline 
celery & Vegetables Celery Artificial Vegetable Model Fakesimulation Faux Veggies Decorative Decoration Pretendkitchen Realistic & 0 & 2 & 0 \\ \hline 
buttermilk & All Natural Buttermilk Pancake and Waffle Mix, Bundle- 32 Once (16 oz 2 pack) & 0 & 2 & 0 \\ \hline 
cakepops & White Cake Pop Sticks, 4-Inch, 100-Count & 0 & 2 & 0 \\ \hline
computer paper & Business Source Premium Multipurpose Copy Paper, 8.5" x 11", 92 Bright, 20 Lb, 2,500 Sheets/5 Reams & 2 & 2 & 0 \\ \hline
bake & OREO Fudge Covered Chocolate Sandwich Cookies, Mothers Day Chocolate, 7.9 oz & 0 & 0 & 2 \\ \hline
\end{tabular}
\caption{Comparison of Model Predictions with Human Labels}
\label{tab:model_comparison}
\end{table}

\textbf{Observations from Model Predictions:}
The experimental analysis conducted sought to ascertain the performance efficacy of GPT-4 and LLAMA2 models in the context of e-commerce search query relevance. A curated dataset was utilized, encompassing a spectrum of queries alongside their corresponding product titles. Each entry was meticulously labeled by human evaluators to reflect the degree of relevance, which ranged from 0 (Completely Irrelevant) to 2 (Highly Relevant). This labeling served as the ground truth against which the model predictions were gauged. Our findings, summarized in Table \ref{tab:model_comparison}, indicate a complex landscape where model predictions vary in alignment with human judgment.

\begin{itemize}
    \item The LLAMA2 model exhibited high alignment with human relevance labels, accurately predicting the relevance for items such as ``McKesson Medical Tape" and ``AHA Sparkling Water," where the query and product titles directly corresponded.
    \item GPT-4 showed a propensity for overgeneralization, evident in its classification of ``shredded chicken" as highly relevant to a ``Chicken Shredder," suggesting a reliance on keyword matching rather than nuanced understanding.
    \item LLAMA2 demonstrated robust discernment capabilities, correctly identifying indirect relevance in queries such as ``glass cleaner" and ``dairy free cheese," which were misclassified by GPT-4, underscoring the necessity for context-aware models in e-commerce search.
\end{itemize}

\textbf{Analysis of Model Performance:}
The comparative analysis reveals both the potential and limitations of current LLMs in discerning product relevance within e-commerce searches. It underscores the need for models to evolve beyond mere keyword matching towards a more sophisticated interpretation of user intent and context. Notably, LLAMA2's conservative approach in assigning relevance could be indicative of its refined understanding of query specificity, although it occasionally results in underestimating relevance where a human evaluator might infer a broader intent.

\textbf{Efficiency and Security Advantages:}
Our fine-tuned LLaMa2 model significantly enhances both the efficiency and security of e-commerce search relevance tasks compared to  GPT-4. Its architecture, optimized for quicker processing, enables real-time application capabilities while ensuring a higher level of data privacy and security. This represents a substantial advancement for e-commerce platforms prioritizing fast and secure search functionalities.


\textbf{Implications for E-commerce Platforms:}
The comparative strengths of LLAMA2 in our study hold profound implications for e-commerce platforms. As online shopping continues to grow, the ability of platforms to deliver accurate and relevant search results directly impacts customer satisfaction and, consequently, business outcomes. Sophisticated NLP solutions like LLAMA2 can transform user experiences by more accurately interpreting and responding to user queries, thus increasing the likelihood of customer retention and repeat purchases.

The effectiveness of LLAMA2 in understanding nuanced user intents and delivering contextually appropriate product suggestions can lead to a more personalized shopping experience. This personalization is crucial for converting searches into sales, as customers are more likely to purchase products that closely match their search intentions. Furthermore, improved search relevancy reduces the customer's effort to find the right product, enhancing their overall shopping experience and increasing the chances of customer loyalty.

%% file: 060conclusion.tex
In this paper, we presented a novel approach to enhancing e-commerce search relevance by integrating Low-Rank Adaptation (LoRA) with the LLAMA2 7B model. Our methodology focused on fine-tuning a large language model specifically for the nuanced requirements of sponsored product search in an e-commerce environment, particularly Walmart.com.

The results of our experiments indicate a significant improvement in search relevance accuracy. With an accuracy of \testAccuracy on our test set, the adapted LLAMA2 7B model outperformed traditional models and established a new standard for e-commerce search relevance. This performance demonstrates the model's enhanced capability to accurately classify and understand complex consumer queries, thereby significantly improving the online shopping experience.


While our evaluation primarily focused on sponsored search, the approach we developed is equally effective for broader e-commerce search contexts. This adaptability highlights the practical utility and versatility of our method across various e-commerce search scenarios.

Despite these advancements, there remain areas for further exploration. Future work could involve extending the model's applicability to more diverse datasets, exploring other domains where such a model could be beneficial, and continuously refining the model to adapt to the evolving nature of consumer language and preferences.

In conclusion, this study offers a significant contribution to the fields of natural language processing and e-commerce search engines. By leveraging the power of advanced language models like LLAMA2 7B and the efficiency of techniques like LoRA, we have opened new pathways for enhancing user interaction and satisfaction in online retail platforms. This work not only enriches the academic domain but also provides practical, impactful solutions for the e-commerce industry.

%% file: BIB/rokon.bib
@inproceedings{aiello2016role,
  title={The role of relevance in sponsored search},
  author={Aiello, Luca and Arapakis, Ioannis and Baeza-Yates, Ricardo and Bai, Xiao and Barbieri, Nicola and Mantrach, Amin and Silvestri, Fabrizio},
  booktitle={Proceedings of the 25th ACM International on Conference on Information and Knowledge Management},
  pages={185--194},
  year={2016}
}

@book{abhishek2019advertising,
  title={Advertising on online marketplaces: Information asymmetry and the relevance of sponsored listings},
  author={Abhishek, Vibhanshu},
  year={2019},
  publisher={SSRN}
}

@inproceedings{rao2020product,
  title={Product insights: Analyzing product intents in web search},
  author={Rao, Nikitha and Bansal, Chetan and Mukherjee, Subhabrata and Maddila, Chandra},
  booktitle={Proceedings of the 29th ACM International Conference on Information \& Knowledge Management},
  pages={2189--2192},
  year={2020}
}

@inproceedings{su2018user,
  title={User intent, behaviour, and perceived satisfaction in product search},
  author={Su, Ning and He, Jiyin and Liu, Yiqun and Zhang, Min and Ma, Shaoping},
  booktitle={Proceedings of the Eleventh ACM International Conference on Web Search and Data Mining},
  pages={547--555},
  year={2018}
}

@inproceedings{Wang2023ClickConversion,
  title={Click-Conversion Multi-Task Model with Position Bias Mitigation for Sponsored Search in eCommerce},
  author={Wang, Y. and Xue, Y. and Liu, B. and Wen, M. and Zhao, W. and Guo, S.},
  booktitle={Proceedings of the 46th International ACM SIGIR Conference on Research and Development in Information Retrieval},
  year={2023},
  organization={ACM},
  url={https://dl.acm.org/doi/abs/10.1145/3539618.3591963}
}

@inproceedings{Li2023CommunicativeMARL,
  title={Communicative MARL-based Relevance Discerning Network for Repetition-Aware Recommendation},
  author={Li, K. and Wang, P. and Wang, H. and Liu, Q. and Wang, X. and Wang, D.},
  booktitle={Proceedings of the ACM Web Conference 2023},
  year={2023},
  organization={ACM},
  url={https://dl.acm.org/doi/abs/10.1145/3543507.3583459}
}

@article{hu2021lora,
  title={Lora: Low-rank adaptation of large language models},
  author={Hu, Edward J and Shen, Yelong and Wallis, Phillip and Allen-Zhu, Zeyuan and Li, Yuanzhi and Wang, Shean and Wang, Lu and Chen, Weizhu},
  journal={arXiv preprint arXiv:2106.09685},
  year={2021}
}

@inproceedings{Aiello2016,
  author = {Aiello, L. and Arapakis, I. and Baeza-Yates, R. and Bai, X. and Barbieri, N. and Mantrach, A. and Silvestri, F.},
  title = {The role of relevance in sponsored search},
  booktitle = {Proceedings of the 25th ACM International on Conference on Information and Knowledge Management},
  year = {2016},
  pages = {185--194}
}

@inproceedings{Wang2023,
  author = {Wang, Y. and Xue, Y. and Liu, B. and Wen, M. and Zhao, W. and Guo, S.},
  title = {Click-conversion multi-task model with position bias mitigation for sponsored search in e-commerce},
  booktitle = {Proceedings of the 46th International ACM SIGIR Conference on Research and Development in Information Retrieval},
  year = {2023},
  url = {https://dl.acm.org/doi/abs/10.1145/3539618.3591963}
}

@inproceedings{Rao2020,
  author = {Rao, N. and Bansal, C. and Mukherjee, S. and Maddila, C.},
  title = {Product insights: Analyzing product intents in web search},
  booktitle = {Proceedings of the 29th ACM International Conference on Information \& Knowledge Management},
  year = {2020},
  pages = {2189--2192}
}

@inproceedings{Su2018,
  author = {Su, N. and He, J. and Liu, Y. and Zhang, M. and Ma, S.},
  title = {User intent behaviour and perceived satisfaction in product search},
  booktitle = {Proceedings of the Eleventh ACM International Conference on Web Search and Data Mining},
  year = {2018},
  pages = {547--555}
}

@inproceedings{Li2023,
  author = {Li, K. and Wang, P. and Wang, H. and Liu, Q. and Wang, X. and Wang, D.},
  title = {Communicative marl-based relevance discerning network for repetition-aware recommendation},
  booktitle = {Proceedings of the ACM Web Conference},
  year = {2023},
  url = {https://dl.acm.org/doi/abs/10.1145/3543507.3583459}
}

@article{touvron2023llama,
  title={Llama 2: Open foundation and fine-tuned chat models},
  author={Touvron, Hugo and Martin, Louis and Stone, Kevin and Albert, Peter and Almahairi, Amjad and Babaei, Yasmine and Bashlykov, Nikolay and Batra, Soumya and Bhargava, Prajjwal and Bhosale, Shruti and others},
  journal={arXiv preprint arXiv:2307.09288},
  year={2023}
}

@article{achiam2023gpt,
  title={Gpt-4 technical report},
  author={Achiam, Josh and Adler, Steven and Agarwal, Sandhini and Ahmad, Lama and Akkaya, Ilge and Aleman, Florencia Leoni and Almeida, Diogo and Altenschmidt, Janko and Altman, Sam and Anadkat, Shyamal and others},
  journal={arXiv preprint arXiv:2303.08774},
  year={2023}
}
